\documentclass[twocolumn,superscriptaddress]{revtex4}
\usepackage{graphicx}
\usepackage{subfig}
\usepackage{float}
\usepackage{bm}
\usepackage{color}
\usepackage{amsmath}
\usepackage{amsfonts}
\usepackage{amssymb}
\usepackage{natbib}
\usepackage{latexsym}
\usepackage{mathrsfs}

\def \ee{\end{equation}}
\def \be{\begin{equation}}
\def \eea{\end{eqnarray}}
\def \bea{\begin{eqnarray}}

\begin{document}

\title{Energy extracted from   Hartle-Thorne strange stars}

\author{Alpha O. Djalo}
\affiliation{New Rochelle High School Science Research Program, 265 Clove Rd New Rochelle, New York 10801, USA.} 

\author{Felipe A. Asenjo} \email{felipe.asenjo@uai.cl}%
\affiliation{Facultad de Ingeniería y Ciencias, Universidad Adolfo Ibáñez,  Peñalol\'en 7941169, Santiago, Chile. }%

\date{\today}

\begin{abstract}
It is discussed how Penrose process can be used in a strange star, described by the Hartle-Thorne metric, to extract its rotational energy. This metric has an ergosphere region that depends only on its angular momentum.  It is shown that massive particles with negative energy can exist in this ergosphere, exploring the conditions on the radial distance and  on angular momentum for this to occur. It is also calculated the total amount of rotational energy that can be extracted from the strange star, and how these conditions cannot be fulfilled  by  a  star with regular matter.
\end{abstract}

\maketitle


\section{Introduction}

Ergosphere in a black hole is the region of curved spacetime, outside the event horizon, in which the temporal part of the metric change its sign. This region is a consequence of the the non spherically symmetric form of the spacetime.  

A very interesting feature of ergosphere is that it can be used to extract rotational energy of black holes, as it allows the possibility to have negative energy particles.
This is the so-called Penrose process \cite{penrose,wald}. The most important issue about ergosphere is that they do not belong to the realm of black holes. They are a general and robust characteristic of several rotating spacetime metrics. In particular, those describing some kind of stars, called strange stars.
Thus, one can wonder if   Penrose process can be applied to those stars, to extract their rotational energy in a simpler fashion, using their reachable star's ergosphere. The purpose of this work is to explore this idea for the Hartle-Thorne  metric \cite{hrthon}, which describe strange stars. This metric
is an approximated solution of vacuum Einstein
field equations  describing the exterior of any slowly and rigidly rotating,
stationary and axially symmetric body. 

We will show that Penrose process is feasible for these kind of stars, although the amount of extracted rotational energy is small within the approximations of the metric. 
In the following section, we review this metric and its corresponding ergosphere, to later, use the Penrose process to demostrate that it is achievable to extract energy from this kind of stars. Lasty, we discuss the how the approximations of the Hartle-Thorne metric constrains the type of content composition of the star, being the Penrose process only able to occur in strange stars with a content with no positive energy and pressure.

\section{Ergosphere and Penrose process}

This metric describe a rotating (rigid) star, with its corresponding angular momentum quadrupole moment. This metric is a solution of Einstein equation   up
to the second order  in the  angular momentum of the star, and up to the first order in
its quadrupole moment. In this case, this metric is given in spherical coordinates as the form $ds^2=g_{tt} dt^2+g_{rr} dr^2+g_{\theta\theta} d\theta^2+g_{\phi\phi}d\phi^2+2 g_{t\phi} dtd\phi$, where ($c=1=G$)
\cite{metric} 
\begin{eqnarray}
g_{tt}&=&\left(1-\frac{2M}{r}\right)\left(1+j^2 F_1+q F_2 \right)\, ,\nonumber\\
g_{rr}&=& -\left(1-\frac{2M}{r}\right)^{-1}\left(1+j^2 G_1-q F_2 \right)\, ,\nonumber\\
g_{\theta\theta}&=&-r^2 \left(1+j^2 H_1+q H_2\right)\, ,\nonumber\\
g_{\phi\phi}&=& g_{\theta\theta} \sin^2\theta\, ,\nonumber\\
g_{t\phi}&=&-\frac{2M^2 j}{r}\sin^2\theta\, ,
\label{metricomponents}
\end{eqnarray}
where $r$ is the radial coordinate from the center of the strange star, and $M$ is its mass. Besides, $j$ and $q$ are the dimensionless angular momentum and dimensionless quadropole of the star. The other functions $F_1$, $F_2$, $G_1$, $H_1$ and $H_2$ depend on the radial coordinate and the polar angle $\theta$. They are detailed in the Appendix.

From the above metric, we can calculate the event horizon. This is located at  $r_H$, which is obtained when the condition
$g_{tt}g_{\phi\phi}-g_{t\phi}^2=0$ is fulfilled. At
to the second order  in the  angular momentum  and at  first order in
the quadrupole moment, we obtain
\begin{eqnarray}
    r_H=2M\left(1-\frac{j^2}{16}\left(1+7\cos^2\theta\right)+\frac{5q}{16}\left(1-3\cos^2\theta\right)\right)\, .
    \label{horzionr}
\end{eqnarray}

The Hartle-Thorne metric share similar features to the Kerr metric. One of this is the 
 existence of a region in which the time component of the metric changes its sign outside the event horizon. This is the ergosphere and its the outer limit $r_E$ is located at when condition $g_{tt}=0$ is fulfilled. In this case, we obtain
\begin{eqnarray}
    {r_E}={2M}\left(1+\frac{j^2}{16}\left(3-11\cos^2\theta\right)+\frac{5q}{16}\left(1-3\cos^2\theta\right)\right)\, .
    \label{horzionr2}
\end{eqnarray}

We can readily see that $r_E> r_H$. 
Therefore, the ergosphere  of a strange star described by the Hartle-Thorne metric is the spacetime region, outside the even horizon of the star, in between \eqref{horzionr} and \eqref{horzionr2}
\begin{eqnarray}
    r_E-r_H=\frac{M j^2}{2}\sin^2\theta\, ,
\end{eqnarray}
depending only on the angular momentum of the star. This region always exist for this metric, vanishing at the star's poles, and being maximum at its equatorial plane.

Because the existence of the ergosphere is allowed in this metric, the Penrose process can be used to   extract rotational energy from this star. Similar to the Kerr metric, the Killing vector $\xi_\mu$ of this metric (which becomes the time translation at infinity) is defined through $\xi_\mu \xi^\mu=g_{tt}$. Inside the ergosphere, thus, it becomes spacelike. Therefore, a particle with energy $E=\xi_\mu p^\mu$ (where $p^\mu$ is the particle four-momentum) can acquire a  negative energy inside the ergosphere. The Penrose process applied to the Hartle-Thorne strange stars would work in a similar fashion to the one in
Kerr metric. A positive-energy particle, free-falling from infinity,  breaks in two parts within the ergosphere. If one part acquires negative energy and falls crossing the event horizon, the other part could return to infinity with larger energy compared to the initial one. Thus, the negative-energy particle that falls in the star produces a diminishing in its angular momentum, allowing to decrease the star rotation.

In order to explore this possibility, let us calculate the energy of a massive particle falling to this star. We start by noticing that, in the star's ergosphere, any massive particle must rotate, as no radial falling is allowed. This can be understood from the radial falling velocity of photons, obtained from $ds^2=0$ with metric \eqref{metricomponents}. This radial velocity for light is
${dr}/{dt}=\left(-{g_{tt}}/{g_{rr}}\right)^{-1/2}
$, which becomes imaginary inside the ergosphere. Therefore, no radial motion (for massless or massive particles) is possible within the ergosphere. Light then move only in co-rotational or counter-rotational trajectories in the ergosphere. These trajectories are obtained from $ds^2=0$, when $dr=0=d\theta$, allowing us to obatain the tangential light velocity $v_{l\pm}$ of those trajectories
\begin{equation}
    v_{l\pm}=R_\phi\frac{g_{t\phi}}{g_{\phi\phi}}\pm \frac{R_\phi}{g_{\phi\phi}}\sqrt{g_{t\phi}^2-g_{\phi\phi}g_{tt}}\, ,
\end{equation}
where $v_{l+}$ ($v_{l-}$) is the velocity of the co-rotational (counter-rotational) light motion with respect to the direction of rotation of the star. Here, 
$R_\phi=\sqrt{g_{\phi\phi}}$
is the effective circumference radius of the orbit.

Similarly, we can calculate the tangential velocity of the trajectories of massive particles inside the ergosphere (it cannot be larger than the co-rotational tangential light velocity). From $E=\xi_\mu p^\mu$, we obtain that the energy per mass unit $m$ is given by
\begin{equation}
    \frac{E}{m}=g_{tt} \left(\frac{dt}{ds}\right)+g_{t\phi} \left(\frac{d\phi}{ds}\right)\, .
\end{equation}
This energy can be positive or negative in the ergosphere. Therefore, we can obtain when it vanishes. This gives a condition on the massive particle velocity for trajectories with zero energy within the ergosphere
\begin{equation}
  v_m=R_\phi\frac{g_{tt}}{g_{t\phi}}\, . 
  \label{massivetangetial}
\end{equation}
Thus, massive particles with positive energy $E>0$ in the ergosphere will have tangential velocities larger than $v_m$. Otherwise, if they have negative energy $E<0$, their tangential velocity should be less than $v_m$. 
However, as the motion of massive particles is bounded by the speed of light, then the only possibility for the existence of massive particle trajectories with negative energies in the ergosphere are those with velocities $v$  smaller than velocity $v_m$
and larger than the counter-rotating tangential light velocity, $v_{l-}<v<v_m$.

\begin{figure}
\includegraphics[width = 3.4in]{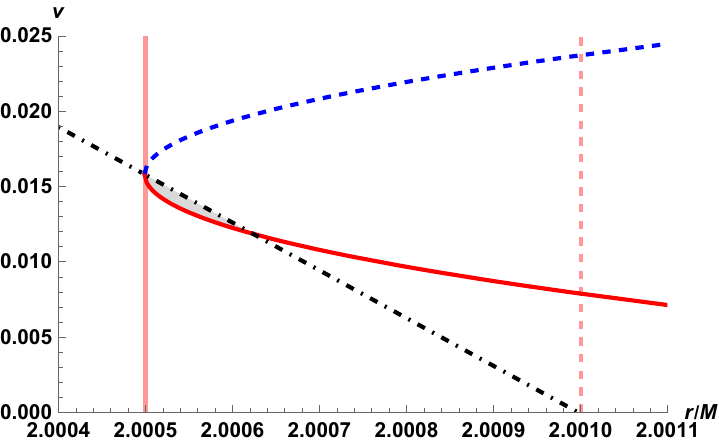}
\caption{Radial dependence of tangential velocities for light (solid red curve for counter-rotating and dashed blue curve for co-rotating) and massive particles with zero energy (black dot-dashed curve) within the ergosphere (region between vertical lines). The gray area shows the velocities that allow massive particles to have negative energies inside the ergosphere.}
\label{figura1}
\end{figure}

In order to show that this condition is fulfilled for this metric, and the existence of negative energy particles in strange stars,  the tangential velocities of light and massive particles in the ergosphere is depicted in Fig.~\ref{figura1} for the arbitrary  case of $j^2=10^{-3}=q$, when $\theta=\pi/2$ (where the ergosphere region is maximum), in terms of normalized radial distance $r/M$. In the figure is shown the radial behavior of these velocities.  In vertical solid line is shown the  event horizon radius \eqref{horzionr} for these values of $j^2$ and $q$. Similarly, in vertical dashed line, the ergosphere outer radius \eqref{horzionr2} is shown
for the same values of $j^2$ and $q$. The region between these two vertical lines is the ergosphere. The radial dependence of the counter-rotating tangential light velocity $v_{l-}$ is shown as a solid red curve, while the co-rotating tangential light velocity $v_{l+}$ appears as the dashed blue curve. Lastly, the tangential velocity $v_m$ for massive particles with zero energy, given in Eq.~\eqref{massivetangetial}, is shown as a black dot-dashed curve.
Massive particle  moving in the ergosphere with negative energies will present tangential velocities with values below the black dot-dashed curve. Therefore, the only possibility for the existence of massive particle trajectories with negative energies in the ergosphere are those with velocities 
 above the counter-rotating tangential light velocity. This is shown as a gray area in Fig.~\ref{figura1}. The existence of this area show the possibility of  extraction
of rotational energy from the star using negative energy particles that could cross the event horizon. This area depends on $j$ and $q$, and it decreases for smaller values of $j^2$ and $q$.

\begin{figure}
\includegraphics[width = 3.2in]{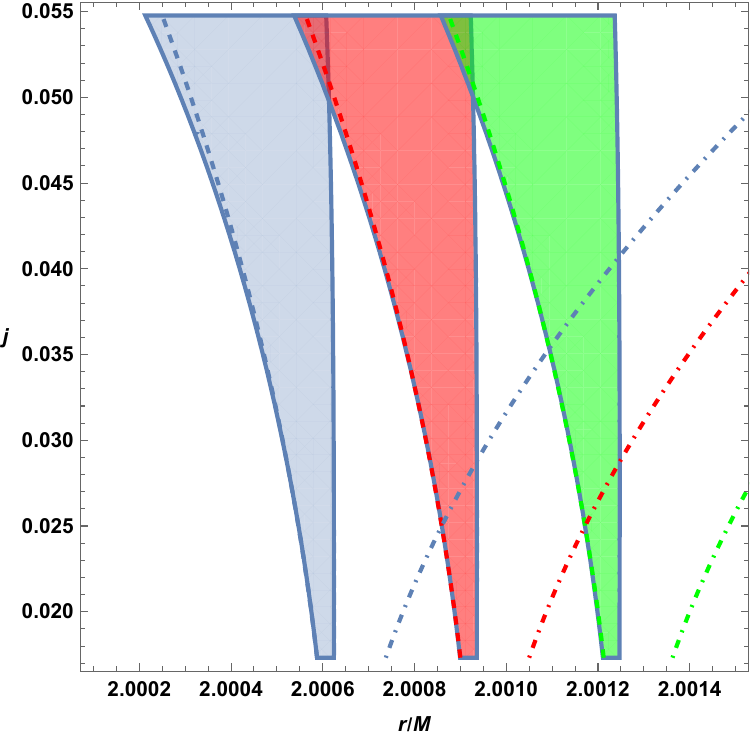}
\caption{Region where $v_m>v_{l-}$ in terms of $r/M$ and $j$, for  $q=10^{-3}$ (blue),  
$q=1.5\times 10^{-3}$ (red), and $q=2\times 10^{-3}$ (green). The event horizon (dashed curves) and outer ergosphere limit (dot-dashed curves) are shown for different values of $q$ (corresponding colors).  }
\label{figura2}
\end{figure}

In general, massive particles with negative energy will be possible in the velocity-region when $v_m>v_{l-}$ . This is shown in Fig.~\ref{figura2}, where the  region in which this condition is fulfilled   is shown in terms of the normalized distance $r/M$ and normalized angular moment $j$ for $q=10^{-3}$ (blue), for 
$q=1.5\times 10^{-3}$ (red), and for $q=2\times 10^{-3}$ (green). The event horizon radius \eqref{horzionr} is depicted by a colored dashed curves,  for the corresponding values of $q$. Similarly, the outer ergosphere radius \eqref{horzionr2} is shown as a dot-dashed curves.
As $q$ increases, the region in which $v_m>v_{l-}$ is satisfied shifts to larger values of  distances $r$, always inside the ergosphere region. Fig.~\ref{figura2} shows that negative energy particles can exist in the $v_m>v_{l-}$ region for a large range of values of $j$ and $q$.

Furthermore, as it always exists the possibilities to have nagative energies particles in the ergosphere, we can calculate the amount of extracted rotational  energy  of the strange star. The angular velocity of the massive particle rotating at distance $r$, as observed
at infinity, is \cite{metric}
\begin{equation}
    \Omega(r)=-\frac{g_{t\phi}}{g_{\phi\phi}}\approx \frac{2M^2 j}{r^3}\, ,
\end{equation}
being independent of the quadrupole moment in the approximations on the Hartle-Thorne metric. When the particle enters the strange star's horizon, its energy $E$ and angular momentum $L$ satisfies the relation $L< E/\Omega(r_H)$ \cite{wald}, where $\Omega (r_H)\approx j/(4M)$, using  Eq.~\eqref{horzionr}. This  implies that negative energy particles carry negative angular momentum.
Then, the change of mass and angular momentum of the star fulfills $\delta J < \delta M/\Omega(r_H)$, which can be put as $\delta M_{\mbox{\footnotesize{irr}}}>0$, where, for our case, the irreducible mass is
\begin{equation}
M_{\mbox{\footnotesize{irr}}}=M \left(1-\frac{j^2}{8}\right)<M\, .
\end{equation}
This value  for $M_{\mbox{\footnotesize{irr}}}$ agrees with the small angular momentum limit of the irreducible mass of a Kerr black hole \cite{wald}. Therefore, the Penrose process for strange stars allow to extract a total rotational energy given by
\begin{equation}
    \frac{M-M_{\mbox{\footnotesize{irr}}}}{M}=\frac{j^2}{8}\, ,
\end{equation}
which is a small, but non-vanishing, amount of energy under the metric approximations.

\section{Remark on the star radius}

The above results shows the possibility of extract
energy from strange stars described by the Hartle-Thorne metric. 
In order to have the physical realistic
scenario described above, a spherical strange star must have a radius $R_{ss}$ which must lie
in between the event horizon and ergosphere radii, 
\begin{eqnarray}
    r_H<R_{ss}<r_E\, .
    \label{radiusineq}
\end{eqnarray} 
This implies that this radius must be in a very narrow set of values of distances allowing the Penrose process to occur outside the star.

A normal non-rotating spherical matter star (having positive energy density and pressure) with mass $M$ must satisfy the Buchdahl's bound \cite{Buchdahl}, which establish that its minimum possible radius is $9M/4$. This is impossible to be fulfilled by inequality \eqref{radiusineq} for the small values required for $j$ and $q$. Even more, for  slowly rotating compact
objects, the minimum radius shows a very small correction  at second order in $j$ \cite{Klenk, Chakraborty}. Therefore, the Penrose process in strange stars described by metric \eqref{metricomponents} requires that the star is not formed by normal matter. Those stars should be composed in a   more exotic way, such as for example dark stars,  gravastars \cite{motola} or boson stars \cite{alfredo}, etc. 

The new physical insights that the Penrose process brings to the dynamics of those stars is a future subject of studies.

\section*{Acknowledgments}

FAA thanks to FONDECYT grant No. 1230094 that partially supported this work.

\appendix{}
\section{Metric functions}

The functions $F_1$, $F_2$, $G_1$, $H_1$ and $H_2$ used in \eqref{metricomponents} are
\begin{eqnarray}
           F_1&=&\left[8M r^4(r-2M)\right]^{-1}\nonumber \\
    &&\left[u^2(48M^6 -8M^5r -24M^4r^2 - 30M^3 r^3\right.\nonumber\\
    &&\left.- 60M^2r^4 + 135Mr^5 - 45r^6)\right.\nonumber\\
    &&\left.+(r-M)(16M^5 + 8M^4r - 10M^2r^3\right.\nonumber\\
    &&\left.- 30Mr^4 + 15r5)\right]+A_1(r)\, ,\nonumber\\
    F_2&=&\left[8M r (r -2M)\right]^{-1}\nonumber\\
    &&\left(5(3u^2-1)(r-M)(2M^2+6M r-3r^2)\right)\nonumber\\
    &&-A_1\, ,\nonumber\\
    G_1&=& \left(8 Mr^4(r-2M)\right)^{-1}\left(L-72M^5 r-3u^2(L-56 M^5 r)\right)\nonumber\\
    &&-A_1\, ,\nonumber\\
H_1&=&\left(8 Mr^4\right)^{-1}(1-3u^2)\left(16M^5+8 M^4 r-10 M^2 r^3\right.\nonumber\\
&&\left.+15 M r^4+15 r^5\right)+A_2\, ,\nonumber\\
H_2&=&\left(8 Mr\right)^{-1}\left(5(1-3u^2)(2M^2-3M r-3r^2)\right)-A_2\, ,
\end{eqnarray}
where $u=\cos\theta$, and
\begin{eqnarray}
    A_1&=&\frac{15 r}{16M^2}(r-2M)(1-3u^2)\nonumber\\
    &&\ln(r/(r-2M))\, ,\nonumber\\
     A_2&=&\frac{15 }{16M^2}(r^2-2M^2)(3u^2-1)\nonumber\\
    &&\ln(r/(r-2M))\, ,\nonumber\\
    L&=&    80 M^6 + 8M^4r^2 + 10M^3r^3 + 20M^2r^4 \nonumber\\
    &&- 45Mr^5 + 15r^6\, .
\end{eqnarray}

\end{document}